\documentclass[prd,aps,floats,amssymb,preprint,nofootinbib]{revtex4}

\usepackage{epsfig}
\usepackage{amssymb}
\usepackage{bbm}
\usepackage{color}
\usepackage{ulem}

\usepackage{enumitem}
\usepackage{graphicx}
\usepackage{subfigure}
\usepackage{amsmath}
\usepackage{bbm}
\usepackage{color}

\newcommand{\be}{\begin{equation}}
\newcommand{\ee}{\end{equation}}
\newcommand{\bea}{\begin{eqnarray}}
\newcommand{\eea}{\end{eqnarray}}
\newcommand{\beas}{\begin{eqnarray*}}
\newcommand{\eeas}{\end{eqnarray*}}
\newcommand{\nn}{\nonumber}

\def\mn{_{\mu \nu}}

\begin{document}


\title{Stability and superluminality of spherical DBI galileon solutions}
\author{Garrett L. Goon\footnote{ggoon@sas.upenn.edu}, Kurt Hinterbichler\footnote{kurthi@physics.upenn.edu} and Mark Trodden\footnote{trodden@physics.upenn.edu}}

\affiliation{Center for Particle Cosmology, Department of Physics and Astronomy, University of Pennsylvania,
Philadelphia, Pennsylvania 19104, USA}

\date{\today}

\begin{abstract}

 The DBI galileons are a generalization of the galileon terms, which extend the internal galilean symmetry to an internal relativistic symmetry, and can also be thought of as generalizations of DBI which yield second order field equations.  We show that, when considered as local modifications to gravity, such as in the Solar system, there exists
 a region of parameter space in which spherically symmetric static solutions exist and are stable.   However, these solutions always exhibit superluminality, casting doubt on the existence of a standard Lorentz invariant UV completion.
 
\end{abstract}

\maketitle

\setcounter{footnote}{0}

\section{Introduction}
The Dirac-Born-Infeld (DBI) action, which describes the dynamics of a brane embedded in a higher dimensional spacetime, has provided an 
important setting within which to study inflation~\cite{Silverstein:2003hf,Alishahiha:2004eh}, late-time cosmic acceleration~\cite{Ahn:2009xd}, tunneling~\cite{Brown:2007zzh}, and exotic topological defects~\cite{Andrews:2010eh,Babichev:2006cy,Sarangi:2007mj,Bazeia:2007df,Babichev:2008qv}. The DBI action has 
been extensively studied in recent years, and its rather special properties are now well-understood.

At the same time, increasing attention has been paid to induced gravity theories, such as the Dvali-Gabadadze-Poratti (DGP) model~\cite{Dvali:2000hr}.  In these,
branes in extra dimensions again form the basic objects, but the Einstein-Hilbert action for gravity is written both in the bulk and on the the
branes themselves, leading to a highly nontrivial behavior of the resulting $4$-dimensional effective theory on the brane.  The theory admits a limit which contains a scalar field $\pi$ which interacts through a higher derivative cubic coupling possessing an internal galilean invariance $\delta\pi=\omega_{\mu}x^{\mu} +\epsilon$ (with $\omega_{\mu}$ and
$\epsilon$ infinitesimal constants) and second order field equations~\cite{Luty:2003vm,Nicolis:2004qq}.  These general properties can be generalized to higher order interactions, and the resulting theories are know as galileons~\cite{Nicolis:2008in}.  

It has recently been shown~\cite{deRham:2010eu} that the galileon and DBI theories are intimately related.  Beginning from a co-dimension one probe brane in a 5D Poincare symmetric bulk, a 4D action can be formed using solely the 4D Lovelock invariants and the boundary terms of 5D Lovelock invariants.  The resuting action then consists of \textit{DBI galileon} terms.  These terms are generalizations of the square root DBI action, in the sense that they share its symmetries and yield second order field equations.  In a small field limit,  the relativistic symmetry stemming from the 5D Poincare symmetry and brane reparametrization invariance reduces to the galilean symmetry of the DGP model and the DBI galilean terms reduce to the galilean terms catalogued in~\cite{Nicolis:2008in}.  Several authors~\cite{Deffayet:2010zh,Hinterbichler:2010xn,Padilla:2010ir,Padilla:2010de} have now demonstrated that a natural generalization of these galileon models to co-dimension greater than one exists, and that many features, including a powerful non-renormalization theorem and a consistent effective field theory~\cite{Hinterbichler:2010xn}, hold in general.

In this paper we find general static spherically symmetric solutions of DBI galileon theories, and explore their stability.  Such an analysis was performed for the ordinary galileons in~\cite{Nicolis:2008in}, and for multi-galileon theories in~\cite{Andrews:2010km,Padilla:2010tj}. In the case of the DGP model~\cite{Nicolis:2008in}, it was found that for some choices of parameters, stable solutions exists but always contain superluminal signal propagation.  We follow the same approach here, extending the results to the DBI galileons, and reach similar conclusions.  The analysis is only valid in the $M_p\to \infty$ limit.  As shown in \cite{Deffayet:2010qz}, the stability of these theories depends on terms suppressed by the square of the Planck mass.

\section{DBI galilean terms and equations of motion}

We are interested in the generalizations of DBI discussed in~\cite{deRham:2010eu}.  The relevant theory consists of a single scalar $\pi$, in $3+1$
dimensions, with an action invariant under the internal relativistic symmetry
\be
\label{poincaresymmetry} 
\delta \pi=\omega_{\mu}x^\mu-\omega^\mu\pi\partial_\mu\pi+\epsilon \ ,
\ee 
with $x^\mu$ the spacetime coordinate, $\omega_{\mu}$ a constant infinitesimal vector, and $\epsilon$ an infinitesimal constant. 

To construct the action for $\pi$, one follows the prescription of~\cite{deRham:2010eu} (see also~\cite{Hinterbichler:2010xn} for further details).  
Consider an embedding of a 3-brane in flat 5-d Minkowski space, $X^A(x^\mu)$ (where $A$ is the 5-d bulk index and $\mu$ the 4-d world-volume index), and a world-volume action which is invariant under world-volume  reparametrizations and bulk Poincare transformations.  The reparametrization invariance forces the action to be a diffeomorphism scalar constructed out of the induced metric $g_{\mu\nu}(x)\equiv {\partial X^A\over\partial x^\mu} {\partial X^B\over\partial x^\nu} G_{AB}\left(X(x)\right)$, where $G_{AB}$ is the bulk metric as a function of the embedding variables $X^A(x)$.  Poincare invariance requires the bulk metric to be the flat Minkowski metric $G_{AB}(X)=\eta_{AB}$.  We then fix the gauge $X^\mu(x)=x^\mu$, and let the unfixed degree of freedom be $X^5\equiv \pi$, so that the induced metric becomes
\be
 g_{\mu\nu}=\eta_{\mu\nu}+\partial_\mu \pi\partial_\nu\pi \ .
\ee
Any action which is a diffeomorphism scalar, evaluated on this metric, will yield an action for $\pi$ having the 
invariance~(\ref{poincaresymmetry}), in which $\omega_\mu$ is a boost in the fifth direction along with a compensating gauge transformation to maintain the gauge choice $X^\mu(x)=x^\mu$.  The parameter $\epsilon$, the shift on $\pi$, is the translation in the fifth dimension.  In addition, the action will have the usual $4$-dimensional spacetime Poincare invariance, which combines with the boost and the shift (\ref{poincaresymmetry}) to form the full 5-d Poincare group.  

The ingredients available to construct such an action are the induced metric $g_{\mu\nu}$, the covariant derivative $\nabla_\mu$ compatible with the induced metric, the Riemann curvature tensor $R^{\rho}_{\ \sigma\mu\nu}$ corresponding to this derivative, and the extrinsic curvature $K_{\mu\nu}$ of the embedding.  Thus, the most general action is
\be
\label{generalaction} 
S=\left. \int d^4x\ \sqrt{-g}F\left(g_{\mu\nu},\nabla_\mu,R^{\rho}_{\ \sigma\mu\nu},K_{\mu\nu}\right)\right|_{g_{\mu\nu}=\eta_{\mu\nu}+\partial_\mu \pi\partial_\nu\pi} \ ,
\ee

For example, the DBI action arises from
\be  
\int d^4x\ \sqrt{-g}\rightarrow  \int d^4x\ \sqrt{1+(\partial\pi)^2} \ .
\ee
where here, and in the remainder of this paper, we use the mostly plus metric convention.

As detailed in~\cite{deRham:2010eu}, only certain choices of $F$ in (\ref{generalaction}) will lead to theories that have second order equations of motion; the Lovelock invariants and their boundary terms.  The terms available are
\begin{align}
\mathcal{L}_{2}&=-\sqrt{-g} \ ,\\
\mathcal{L}_{3}&=\sqrt{-g}K \ ,\\
\mathcal{L}_{4}&=-\sqrt{-g}\,R \ , \\
\mathcal{L}_{5}&=\dfrac{3}{2}\sqrt{-g}\,\mathcal{K}_{GB} \ ,
\end{align}
where 
\be
\mathcal{K}_{\rm GB}=-\frac 23 K\mn^3+K K\mn^2-\frac 13 K^3-2(R\mn-\frac 12 R g\mn) K^{\mu\nu}
\ee
is the Myers boundary term from the second order Lovelock invariant in the bulk \cite{Myers:1987yn}, and $\mathcal{L}_{3}$ is the Gibbons-Hawking-York boundary term for the Einstein-Hilbert action in the bulk \cite{Gibbons:1976ue,York:1972sj}.

In the following, we use the notation $\Pi$ for the matrix of partials $\Pi_{\mu\nu}\equiv\partial_{\mu}\partial_\nu\pi$, and $[\Pi^n]\equiv Tr(\Pi^n)$, e.g. $[\Pi]=\square\pi$, $[\Pi^2]=\partial_\mu\partial_\nu\pi\partial^\mu\partial^\nu\pi$, as well as $[\pi^n]\equiv \partial\pi\cdot\Pi^{n-2}\cdot\partial\pi$, e.g. $[\pi^2]=\partial_\mu\pi\partial^\mu\pi$, $[\pi^3]=\partial_\mu\pi\partial^\mu\partial^\nu\pi\partial_\nu\pi$.  Indices are raised and lowered with the flat metric, and we use the mostly plus signature.  Also, we define
\be 
\gamma={1\over \sqrt{1+(\partial\pi)^2}} \ .
\ee

In terms of these combinations of the field $\pi$ and its derivatives, the terms above become
\begin{align}
\mathcal{L}_{2}&=-\sqrt{1+(\partial \pi)^{2}} \ , \\
\mathcal{L}_{3}&=-\left [\Pi\right ]+\gamma^{2}\left [\pi^3\right ] \ , \\
\mathcal{L}_{4}& =-\gamma \left (\left [\Pi \right ]^{2} -\left [\Pi^{2}\right ]\right )-2\gamma^{3}\left (\left [\pi^{4}\right ]-\left [\Pi\right ]\left [\pi^3\right ]\right ) \ ,\\
\mathcal{L}_{5}& =-\gamma^{2}\left (\left [\Pi\right ]^{3}+2\left [\Pi^{3}\right ]-3\left [\Pi\right ]\left [\Pi^{2}\right ]\right )-\gamma^{4}\left (6\left [\Pi\right ]\left [\pi^{4}\right ]-6\left [\pi^{5}\right ]-3\left (\left [\Pi\right ]^{2}-\left [\Pi^{2}\right ]\right )\left [\pi^3\right ]\right ) \ ,
\end{align}
where we have explicitly retained all total derivatives.

In $3+1$ dimensions, the above terms are the only ones possessing the symmetry~(\ref{poincaresymmetry}) and yielding second order equations of motion.  The first term is the DBI action, which when expanded gives the standard kinetic term for the scalar.  The second is the relativistic version of the cubic DGP $\pi$-lagrangian (up to a total derivative).  These are the DBI generalizations of the galileons studied in~\cite{Nicolis:2008in}.  The galileons are recovered by expanding in powers of the field $\pi$ and taking the lowest non-trivial contribution from each term~\cite{deRham:2010eu}. 

The resulting equations of motion take the form $ \mathcal{E}_{n}=0$, with $n=2,3,4,5$, and
\begin{align}
\mathcal{E}_{2}&=\gamma\left [\Pi\right ]-\gamma^{3}\left [\pi^3\right ] \ ,  \\
\mathcal{E}_{3}&=\gamma^{2} \left (\left [\Pi \right ]^{2} -\left [\Pi^{2}\right ]\right )+2\gamma^{4}\left (\left [\pi^{4}\right ]-\left [\Pi\right ]\left [\pi^3\right ]\right ) \ , \\
\mathcal{E}_{4}&= \gamma^{3}\left (\left [\Pi\right ]^{3}+2\left [\Pi^{3}\right ]-3\left [\Pi\right ]\left [\Pi^{2}\right ]\right )+\gamma^{5}\left (6\left [\Pi\right ]\left [\pi^{4}\right ]-6\left [\pi^{5}\right ]-3\left (\left [\Pi\right ]^{2}-\left [\Pi^{2}\right ]\right )\left [\pi^3\right ]\right ) \ , \\
\mathcal{E}_{5}&= \gamma^{6}\left (\left [\Pi\right ]^{4}-6\left [\Pi^{2}\right ]\left [\Pi\right ]^{2}+8\left [\Pi\right ]\left [\Pi^{3}\right ]+3\left [\Pi^{2}\right ]^{2}-6\left [\Pi^{4}\right ]\right ) \ .
\label{eom}
\end{align}

These satisfy the following interesting recursion relation noticed in~\cite{deRham:2010eu},
\begin{align}
\dfrac{\delta}{\delta \pi}\left (\sqrt{-g}\, \right )&=K \ , \\
\dfrac{\delta}{\delta \pi}\left (\sqrt{-g}\, K\right )&=R \ , \\
\dfrac{\delta}{\delta \pi}\left (\sqrt{-g}\, R\right )&=\dfrac{3}{2}\mathcal{K}_{GB} \ , \\
\dfrac{\delta}{\delta \pi}\left (\sqrt{-g}\, \mathcal{K}_{GB}\right )&=\dfrac{2}{3}\mathcal{L}_{GB_4} \ ,
\end{align}
where $\mathcal{L}_{GB_4}=R^2-4R\mn^2+R_{\mu\nu\alpha\beta}^2$ is the second order Lovelock invariant.

In this paper we consider a theory containing these terms with arbitrary coefficients $d_n$, and which is linearly coupled to the trace $T$ of the energy momentum tensor of matter, so that the complete Lagrangian density is
\be 
{\cal L}=\sum_{n=2}^5 d_n {\cal L}_n+\pi T \ ,
\ee
with equation of motion ${\cal E} =0$, where
\be 
\label{fullequations} 
{\cal E} \equiv \sum_{n=2}^5 d_n {\cal E}_n+T \ .
\ee

The linear coupling is not invariant under the symmetry operation (\ref{poincaresymmetry}).  Rather, it was chosen for simplicity and for comparison with the results of \cite{Nicolis:2008in} where the same choice was made.  It is also the coupling that arises if the scalar is considered as a modification to gravity that conformally mixes with the graviton, as happens in the DGP model.  Although there may exist physically interesting couplings which obey the symmetry, the simplest example, $\partial_{\mu}\pi\partial_{\nu}\pi T^{\mu\nu}$, which arises naturally from the brane construction, gives no contribution to the equations of motion for static sources.

Our goal is to derive constraints on these models from the requirements of stability and subluminality of mode propagation around
spherically symmetric backgrounds. We shall begin this analysis in the next section, but it is important to note that one constraint can be seen immediately;
\be 
d_2>0 \ ,
\ee
since otherwise the kinetic term will yield a ghost (or will be absent, if we set $d_2=0$).

\section{Spherical solutions}

We search for static spherically symmetric solutions to the equations of motion in spherical polar coordinates $(r,\theta,\phi)$, in the presence of a positive mass delta function source at the origin
\be 
T=-M\delta^3(r), \ \ \ M>0 \ .
\ee 
To evaluate the equations of motion we need find only the non-vanishing elements of $\Pi_{\mu\nu}=\partial_{\mu}\partial_{\nu}\pi-\Gamma^{\alpha}_{\mu\nu}\partial_{\alpha}\pi$. These are $\Pi_{rr}=\pi_{,rr}$, $\Pi_{\theta\theta}=r\pi_{,r}$, and $\Pi_{\phi\phi}=r\sin^{2}\theta\pi_{,r}$. Since the flat metric is diagonal we then have
\bea
\left [\Pi^{n}\right ]&=&\left (\Pi_{rr}\eta^{rr}\right )^{n}+\left (\Pi_{\theta\theta}\eta^{\theta\theta}\right )^{n}+\left (\Pi_{\phi\phi}\eta^{\phi\phi}\right )^{n}=\pi_{,rr}^{n}+\dfrac{2\pi_{,r}^{n}}{r^{n}} \ ,\\
 \left [\pi^{n+2}\right ]&=&\pi^{2}_{,r}\left (\Pi_{rr}\right )^{n}\left (\eta^{rr}\right )^{n+1}=\pi^{2}_{,r}\pi_{,rr}^{n} \ .
 \eea

Using these, the equations of motion~(\ref{fullequations}) become 
\begin{align}
\mathcal{E}_{2}&=\dfrac{1}{r^{2}}\dfrac{d}{dr}\left [r^{3} y \right ] \ ,\\
\mathcal{E}_{3}&=\dfrac{2}{r^{2}}\dfrac{d}{dr}\left [r^{3} y^{2} \right ] \ ,\\
\mathcal{E}_{4}&=\dfrac{2}{r^{2}}\dfrac{d}{dr}\left [r^{3}y^{3}\right ] \ ,\\
\mathcal{E}_5&=0 \ ,
\end{align}
where we have defined 
\be
y\equiv\dfrac{\gamma \pi'}{r} \ . 
\ee
The fifth order term vanishes because our focus on static solutions reduces the problem to a three dimensional one, and the fifth order term is trivial in three dimensions. The remaining equations of motion can be written as a polynomial in $y$ as 
\be 
\label{preintegrate} 
\dfrac{1}{r^{2}}\dfrac{d}{dr}\left [r^{3}P(y)\right ]=M\delta^3(r) \ ,
\ee
with
\be
P(y)\equiv d_2y+2d_3y^{2}+2d_4y^{3} \ .
\ee

Note that the equations of motion are a total $r$-derivative.  This is a consequence of the shift invariance $\pi\rightarrow\pi+c$ of the Lagrangian, which has an associated Noether current $J^{\mu}$, in terms of which the equations of motion take the form $\partial_\mu (-J^\mu)=0$. We may therefore integrate the equations of motion once to obtain
\be 
\label{polyequation} 
P(y)={M\over 4\pi r^3} \ .
\ee

We now study the existence of spherically symmetric solutions, and the resulting constraints on the coefficients $d_2,d_3,d_4$.  Our boundary condition is that $\pi$ approaches a constant as $r\rightarrow \infty$.  
The other boundary condition is fixed by the delta function at the origin.  Focusing on small $r$, (\ref{polyequation}) yields
\be 
{\pi'^3\over (1+\pi'^2)^{3/2}}d_4={M\over 8\pi} \ .
\ee
This determines a finite value for $\pi'$ at the origin, and therefore implies that $\pi$ must also be finite there.  
Since the absolute value of the prefactor in front of $d_4$ on the left hand side is always less than unity, we then obtain the constraint
\be
\label{Mconstraint} 
\left|d_4\right|>{M\over 8\pi} \ .
\ee
This constraint is unique to the DBI action - no such constraint arises in the usual galileon theories.  The fourth-order term dominates at short distances, and its non-linearities render $\pi$ finite at the origin.  In particular therefore, note that there are no spherically symmetric static solutions in the pure DBI model, for which $d_3=d_4=0$.  

As we have demonstrated, $\pi'(r)$ ranges from some finite non-zero value at $r=0$, to zero as $r\rightarrow \infty$ (since $\pi$ itself goes to a constant).  Thus, the variable $y=\gamma \pi'/r$ ranges from infinity to zero as $r$ ranges from zero to infinity (we will see shortly that it does so monotonically).  

As $r$ varies from the origin to infinity, the right hand side of~(\ref{polyequation}) ranges from zero to infinity, so the cubic polynomial on the left must do so as well.   Looking at small $y$, along with the requirement $d_2>0$ for a healthy kinetic term, tells us that $P(y)$ intersects the origin and is monotonically increasing near the origin, and hence that $y$ as a function of $r$ is monotonically decreasing in the same region.   As $y$ gets larger ($r$ smaller, $P(y)$ larger), the solution for $y(r)$ must continue to exist and be smooth, which means that $P(y)$ must not have any of its critical points in the region $y>0$.  Thus $P(y)$ monotonically increases for $y>0$, and hence $y(r)$ is monotonically decreasing for $r>0$.  Looking at the form of $y$, this implies in turn that $\pi'(r)$ is monotonic, ranging from some finite value to zero as $r$ goes from zero to infinity.  Integrating, we see that $\pi(r)$ is monotonic as well.

The condition we have then is
\be 
\label{conditionderiv} 
P'(y)=d_2+4d_3y+6d_4y^{2}>0, \ \ \rm{for}\  y>0 \ .
\ee
Focusing on large $y$ implies that $d_4\geq 0$, so that we can now remove the absolute value sign in~(\ref{Mconstraint}).
We already know that $d_2>0$, from the requirement of a healthy kinetic term, but it is worth pointing out that a direct implication of~(\ref{conditionderiv}), 
applied at small $y$, is that spherical solutions do not exist for a ghost-like theory with $d_2<0$.  Furthermore, we are safe if the minimum of $P'(y)$ occurs above zero, which happens if
\be
|d_3|<\sqrt{{3\over 2}d_2 d_4} \ .
\ee  
Otherwise, the largest root of $P'(y)$ must occur for $y\leq0$, which happens if $d_3\geq 0$.  

In summary, the flat space theory is ghost-free and spherical solutions exist if and only if
\be \label{existenceconstraints}
d_2>0,\ \ \ d_4>{M\over 8\pi},\ \ \ d_3>-\sqrt{{3\over 2}d_2 d_4} \ .
\ee

\section{Stability}

The existence of spherically symmetric solutions is, of course, not sufficient to guarantee viability of the theories in question. The next test is
to examine the stability of these solutions.  To do this, we expand the action in perturbations around the spherical solutions 
\be 
\pi(x)=\pi_0( r)+\varphi(x) \ ,
\ee
and isolate the terms quadratic in $\varphi$.  These terms take the form
\be
{\cal S}_\varphi ={1\over 2} \int \! dt \int d^2\Omega\int _0^\infty r^2 dr\  \left [K_{t} (r) \dot\varphi^2 - K_{r} (r) (\partial_r \varphi)^2  - K_{\Omega} (r) (\partial_\Omega \varphi)^2 \right] \ ,
\label{quadraticradial}
\ee
where overdots denote time derivatives, $(\partial_\Omega \varphi)^2=(\partial_\theta\varphi)^2+{1\over \sin^2\theta}(\partial_\phi\varphi)^2$ is the angular part of $(\vec \nabla \varphi)^2$, and the kinetic coefficients $K$ depend on $r$ through the background radial solution $\pi_0(r)$ and its derivatives. Note that the quadratic action contains only second derivatives acting on the perturbations.  This is because the field equations are second order, despite the fact that the lagrangian is higher derivative, as we mentioned earlier.

In order for the solution to be stable, each $K_i(r)$ ($i=t,\ r,\ \Omega$) must be positive for all $r>0$.  If $K_t$ is negative in some region, then localized excitations will be ghostlike and will carry negative energy.  If either of $K_r$, $K_\Omega$ are negative in some region, then it is possible to find
localized perturbations for which gradients lower the energy of the background solution.  This kind of instability, associated with negative gradient energy for certain classes of fluctuations, is more troublesome than a tachyon-like instability associated with a negative mass squared term or upside down potential.  A tachyon-like instability is, like the Jeans instability, dominated by modes with momenta of order the tachyonic mass scale, which can be parametrically smaller than the UV cutoff, and thus computable within the effective theory. By contrast,
the gradient instability can be due to very short wavelength wave-packets with high momentum.  Thus, this instability
also plagues fluctuations right down to the UV cutoff of the theory, so that quantities such as decay rates are dominated by 
the shortest distances in the theory, and cannot be reliably computed within the effective theory.

To obtain explicit expressions for the functions $K_i(r)$, we expand the equations of motion to linear order in $\varphi$
\be 
\label{fluctuationequation}
{\cal E}[\pi_0 + \varphi] \to \frac{\delta {\cal S}_\varphi}{\delta \varphi} = -K_t (r)\, \ddot \varphi +
\frac{1}{r^2} \partial_r \big(r^2 K_r (r)\, \partial_r \varphi \big) + K_\Omega (r) \,  \partial^2_\Omega \varphi \ ,
\ee
where $\partial^2_\Omega={1\over \sin\theta}{\partial\over \partial\theta}\left(\sin\theta{\partial \over \partial \theta}\right)+{1\over \sin^2\theta}{\partial^2\over \partial\phi^2}$ is the angular part of the laplacian.

We begin with the radial perturbations, and find $K_r$ simply by perturbing the radial equation~(\ref{preintegrate}), using a perturbation that depends only on $ r$
\be 
\delta {\cal E}=\dfrac{1}{r^{2}}\dfrac{d}{dr}\left [r^{3}P'(y)\delta y\right ]=\dfrac{1}{r^{2}}\dfrac{d}{dr}\left [r^{2}P'(y)\gamma^3\varphi'\right] \ .
\ee
From this we read off
\be
K_r(r)=\gamma^3P'(y) \ .
\ee
From~(\ref{conditionderiv}), we then see that if the solution exists, then $K(r)$ is automatically positive, since $\gamma>0$.

Now turn to the angular perturbations.  To find $K_\Omega$, we vary the full equations (\ref{fullequations}), allowing the perturbation to depend only on angular variables, and keeping in mind that the background depends only on $r$.  Using the
following useful expressions
\be 
\delta\left [\Pi^{n}\right ]=\dfrac{n\pi'^{n-1}}{r^{n-1}}\partial^{2}_{\Omega}\varphi \ ,\ \ \ \delta\left [\pi^{n}\right ]=0 \ ,\ \ \ \delta\gamma=0 \ ,
\ee
it is simple to show that
\be 
K_{\Omega}(r)=\dfrac{\gamma}{2r} \dfrac{d}{dr}\left[ r^{2}P'(y)\right] \ .
\ee
Recall that the coefficient $d_5$ does not enter in either $K_r$ or $K_\Omega$, because we are still considering static
configurations, for which the fifth DBI term vanishes.

Lastly, we consider the temporal perturbations.  We find $K_t$ by varying the full equations~(\ref{fullequations}), this time allowing the perturbation to depend only on time.  
Once again, some useful expressions
\be 
\delta\left [\Pi\right ]=-\ddot\varphi \ ,\ \ \ \delta\left [\Pi^n\right ]=0 \ \ (n>1) \ ,\ \ \ \  \delta\left [\pi^{n}\right ]=0 \ ,\ \ \ \delta\gamma=0 \ ,
\ee
allow us to show that
\be 
K_{t}(r)=\dfrac{\gamma}{3r^{2}}\dfrac{d}{dr}\left [r^{3}\left (d_2+6d_3y+18d_4y^{2}+24d_5y^{3}\right )\right ] \ .
\ee
We see that $d_5$ enters here for the first time, since we have deviated, at last, from static equations.  

As we have written them, the functions $K_i(r)$ depend on $\gamma$, $r$, $\dfrac{dy}{dr}$ and $y$.  However, we may eliminate $\dfrac{dy}{dr}$ in favor of $y$ by using the implicit function theorem on the function $F(y,r)=P(y)-\dfrac{M}{4\pi r^{3}}=0$.  This yields
\begin{align} \nn
\dfrac{dy}{dr}&=-\dfrac{\partial_{r}F}{\partial_{y}F}=-\dfrac{3}{r}\frac{P(y)}{P'(y)}.
\end{align}
Substituting this into our expressions for the $K_i(r)$ we obtain
\begin{align}
K_r&=\gamma^{3}\left[{d_2+4d_3y+6d_4y^{2}}\right] \ ,\nn\\ 
K_\Omega&=\gamma\left [\dfrac{d_2^{2}+2d_{2}d_{3}y+\left (4d_{3}^{2}-6d_{2}d_{4}\right )y^{2}}{d_2+4d_3y+6d_4y^{2}}\right] \ , \nn\\ 
K_{t}&=\gamma\left [\dfrac{d_2^{2}+\left (4d_2d_3\right )y^{}+12\left (d_{3}^{2}-d_2d_4\right )y^{2}+24\left (d_3d_4-2d_5d_2\right )y^{3}+12\left (3d_4^{2}-4d_3d_5\right )y^{4}}{d_2+4d_3y+6d_4y^{2}}\right ] \ . 
\label{theKs}
\end{align}
Note that the explicit $r$ dependence has canceled out.

Since the solution spans all positive values of $y$ as $r$ varies from zero to infinity, we require $K_t$ and $K_\Omega$ to be positive for all $y>0$.  The denominators in~(\ref{theKs}) are automatically positive, from~(\ref{conditionderiv}).  Given the constraints (\ref{existenceconstraints}), The numerator in $K_{\Omega}$ is positive for 
$d_3 \ge \sqrt{\frac{3}{2} \, d_2 \, d_4}$  which also ensures that the numerator in $K_t$ is positive provided $d_5\le \frac34 \frac{d_4^2}{d_3}$.

The radial solution therefore exists and is stable if and only if
\be \label{finalconstraints}
d_2>0 \ ,\ \ \ d_4>{M\over 8\pi} \ ,\ \ \ d_3\geq \sqrt{{3\over 2}d_2 d_4} \ ,\ \ \ d_5\le \frac34 \frac{d_4^2}{d_3} \ .
\ee

\section{Propagation speed of fluctuations}

As a final test of the viability of the DBI galileon theories, we consider the propagation speeds of small fluctuations around the stable spherical solutions.  For radially propagating fluctuations this speed is
\be
c_r^2 = \frac{K_r}{K_t} \ .
\ee
At large distances from the source (small $y$), this becomes 
\be
c_r^2 = 1 + 4 \frac{d_3}{d_2} y + {\cal O}(y^{4/3}) > 1 \ ,
\ee
where here and in what follows we express $\gamma$ in terms of $y$ via 
\be 
\gamma=\sqrt{1-r^2y^2}=\sqrt{1-\left(M\over 4\pi P(y)\right)^{2/3}y^2} \ .
\ee
Therefore, given the constraints implied by existence and stability of the solutions, this is always superluminal.

At smaller distances (larger $y$), the speed is
\be c_r^2={3d_4^2\over 3d_4^2-4d_3d_5}\left[1-\left(M\over 8\pi d_4\right)^{2/3}\right]+{\cal O}\left({1\over y}\right),\ee
so the propagation speed is subluminal in this region if
\be \label{subluminalconstraint}
d_5<\dfrac{3d_4^{2}}{4d_3}\left (\dfrac{M}{8\pi d_4}\right )^{2/3} \ .
\ee

\begin{figure}[h!]
\begin{center}
\includegraphics[height=3in]{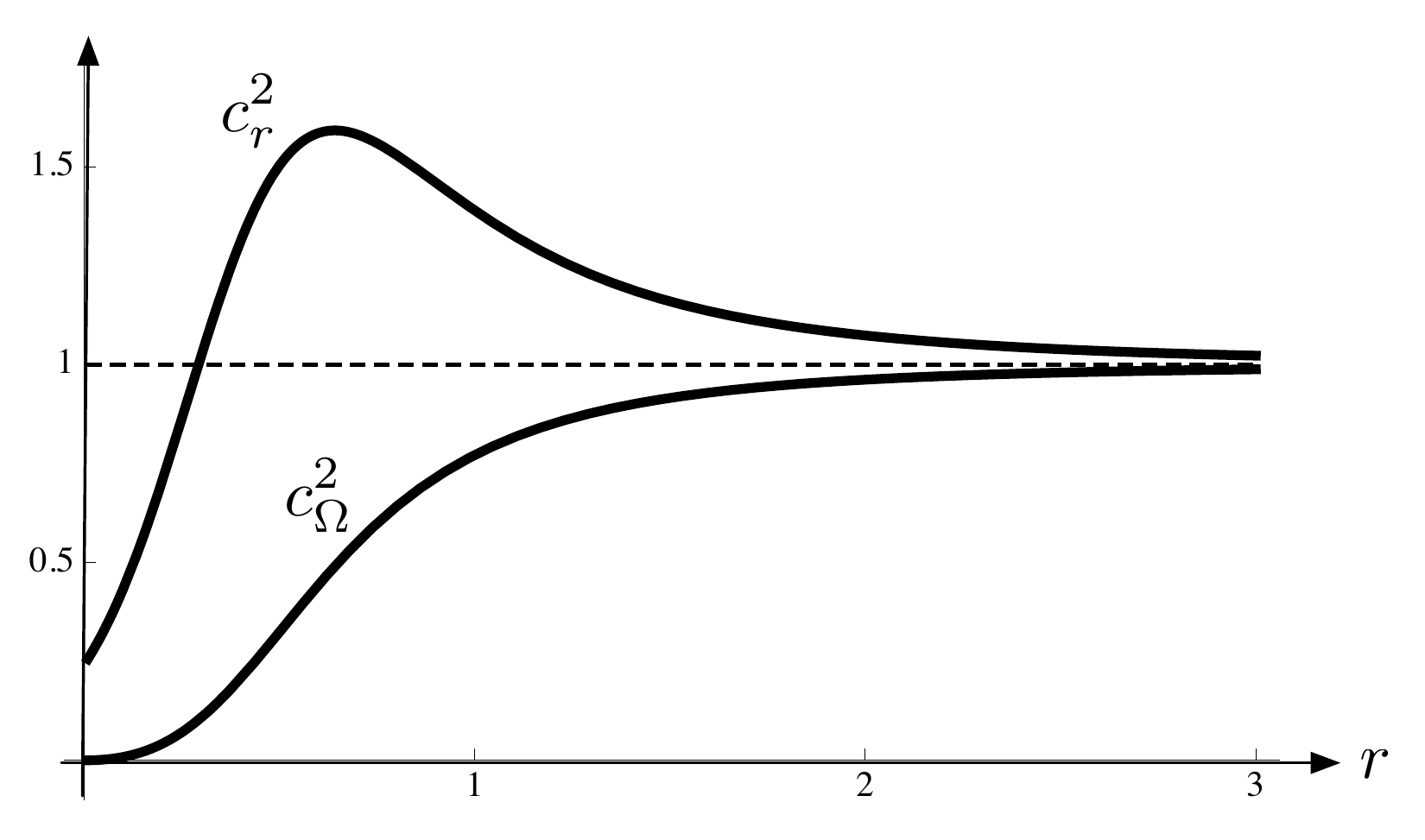}
\caption{Speed of fluctuations $c_r^2$ and $c_\Omega^2$, in the radial and angular directions respectively, for a sample solution satisfying the existence and stability constraints (\ref{finalconstraints}), as well as (\ref{subluminalconstraint}).  The values chosen are $d_2=1$, $d_3=2$, $d_4=1$, $d_5=-1$, $M=1$.}
\label{figure}
\end{center}
\end{figure}

The speed of angular excitations is
\be
c_\Omega^2 = \frac{K_\Omega}{K_t} \ .
\ee
The difference between the numerator and the denominator is, apart from an overall positive factor,
\be
K_\Omega - K_t \sim   -2 d_2d_3y-\left (8d_3^{2}-6d_2d_4\right )y^{2}-24\left (d_3d_4-2d_2d_5\right )y^{3}-12\left (3d_4^{2}-4d_3d_5\right )y^{4} \ .
\ee
Given the constraints (\ref{finalconstraints}), this is always negative, so the speed of angular excitations is always subluminal.  Also, the angular speed goes to zero as $r$ goes to zero.  The radial and angular speeds for a sample solution are shown in figure~\ref{figure}.

Certainly the existence of superluminally propagating modes raises questions about the viability of galileon DBI theories. Whether such a 
feature in really a problem that conclusively rules out a low-energy effective theory is still being debated \cite{Bruneton:2006gf,Babichev:2007dw,Geroch:2010da}, but it has been argued that, 
at the least, it may preclude the possibility of embedding the theory into a local, Lorentz invariant UV completion \cite{Adams:2006sv}.

\section{Conclusions}

The DBI galileon theories establish a natural generalization of, and connection between, the galileon and DBI models through their higher-dimensional realizations and brane actions. In this
paper we have studied spherically symmetric solutions to the DBI galileon models, demonstrating
that there exists a range of parameters in which such solutions exist. We have also examined the stability of these solutions and computed the propagation speeds of perturbations around the solutions. While we have found that there exists a region of parameter space in which our solutions are stable, we have shown that these solutions always exhibit superluminal propagation.  Such behavior is familiar from that of the ordinary galileon theories. Thus, although one might have thought that the $\gamma$ factors appearing for 
DBI galileons could cure the superluminality issues, the results we find here indicate that they do not.  

We have worked in dimensionless units, which corresponds to setting to unity a scale, $\Lambda$, suppressing all the non-linearities in the Lagrangian.  In addition, we have absorbed into the stress tensor a scale, $M_p$, representing the coupling strength.  Restoring these scales, the condition (\ref{Mconstraint}) tells us $d_4\gtrsim M/M_p$, so in gravitational applications, where $M$ is the mass of the Sun and $M_p$ the Planck mass, this tells us that $d_4$ must be huge, of order the solar mass in Planck units.  One might worry that this necessitates strong coupling, but this is not the case because the coefficient $d_2$, which multiplies the kinetic term, may also be chosen to be very large, so that after canonical normalization the true couplings are still small.  

To see the consequences of this, consider expanding the action with the scale $\Lambda$ restored.  The DBI term reads schematically $d_2\Lambda^4\sqrt{1+{(\partial\pi)^2\over \Lambda^4}}\sim (\partial\hat\pi)^2+{1\over d_2\Lambda^4}(\partial\hat\pi)^4+\cdots$, with the canonically normalized field $\hat\pi= d_2^{1/2}\pi$.  The scale suppressing the non-linear terms here is $d_2^{1/4}\Lambda$.  Similarly, the quartic galileon term is, schematically, $d_4\left[1+{(\partial\pi)^2\over \Lambda^4}+\cdots\right]{1\over \Lambda^6}(\partial^2\pi)^2(\partial\pi)^2={d_4\over d_2\Lambda^2}{1\over d_2\Lambda^4}(\partial^2\hat\pi)^2(\partial\hat\pi)^2+{d_4\over d_2\Lambda^2}{1\over d_2^2\Lambda^8}(\partial^2\hat\pi)^2(\partial\hat\pi)^4+\cdots$, which means that the strong coupling scales are $\left(d_2^2\over d_4\right)^{1/6}\Lambda, \left(d_2^3\over d_4\right)^{1/10}\Lambda,\cdots$.  Since $d_4$ is so large, keeping the lowest strong coupling scale reasonably high requires choosing $d_2$ large, say $d_2^2\sim d_4$, in which case all the higher order DBI scales are much higher (corresponding to small coupling), and the theory becomes very similar to the ordinary galileons, explaining why we find conclusions similar to the conclusions in that case.  In addition, note that the coupling to the stress tensor, in terms of the canonically normalized field, is $\sim {1\over d_2^{1/2}M_p}\hat \pi T$, so that the true Planck mass is actually $\sim d_2^{1/2}M_p$, and the necessary size of $d_4$ is actually larger than the solar mass in physical Planck units.

On the other hand, in some situations, it may be too much to demand that the spherical solutions exist for all $r$.  For example, if $\pi$ represents the fifth coordinate of a brane embedding, we should not expect that the brane configuration should be everywhere expressible as a single valued function of the four coordinates $x^\mu$ (the solutions of \cite{Callan:1997kz,Gibbons:1997xz,Gibbons:2001gy,Nastase:2007ut} are examples of this).  In this case, the restrictions on the coefficient $d_4$ may be relaxed.

DBI galileon theories therefore, like the ordinary galileons, face a challenge from the superluminal propagation of perturbations around simple spherically symmetric solutions.  Whether these theories are viable depends on the development of an argument that this superluminality does not lead to the pathologies
that are traditionally associated with this behavior, or whether a modification to the theory or its couplings to matter or gravity can eliminate this behavior.  If the effects of gravity are taken into account, It should be mentioned that the coupling of galileons to gravity is non-trivial if one wishes to keep the equations of motion second order \cite{Deffayet:2009wt,Deffayet:2009mn}, and the issue of superluminality should in principle be re-examined in the full covariant context, though the effects should be Planck suppressed.  It should be straightforward to extend these results to anti-DBI type theories \cite{Mukhanov:2005bu,Babichev:2006vx}, which do not necessarily have a higher dimensional brane interpretation.

\acknowledgments
This work is supported in part by NSF grant PHY-0930521, Department of Energy grant DE-FG05-95ER40893-A020, and by NASA ATP grant NNX08AH27G. MT is also supported by the Fay R. and Eugene L. Langberg chair.

\appendix

\bibliographystyle{apsrev}
\bibliography{DBIgalileonPRD}

\end{document}